# Theoretical Prediction of Heterogeneous Integration of Dissimilar Semiconductor with Various Ultra-Thin Oxides and 2D Materials

*Md Nazmul Hasan, Chenxi Li, Junyu Lai, Jung-Hun Seo\**

Department of Materials Design and Innovation, University at Buffalo, the State University of New York (SUNY), Buffalo, New York, U.S.A, 14260

*Email: junghuns@buffalo.edu

**Abstract**

In this paper, we have built a numerical p-n Si/GaAs heterojunction model using a quantum-mechanical tunneling theory with various quantum tunneling interfacial materials including two-dimensional semiconductors such as hexagonal boron nitride (h-BN) and graphene and ALD-enabled oxide materials such as $HfO_2$, $Al_2O_3$, and $SiO_2$. Their tunneling efficiencies and tunneling current with different thicknesses were systematically calculated and compared. Multiphysics modeling was used with the aforementioned tunneling interfacial materials to analyze changes in strain under different temperature conditions. Considering the transport properties and thermal-induced strain analysis, $Al_2O_3$ among three oxide materials and graphene in 2D materials are favorable material choices that offer the highest heterojunction quality. Overall, our results offer the viable route to guide the selection of quantum tunneling materials for myriad possible combinations of new heterostructures that can be obtained via remote epitaxy and the UO method.

**Introduction**

Semiconductor heterojunctions have long been considered one of the most important building blocks of the semiconductor industry. They have led to various practical electronic and optoelectronic applications such as lighting devices (light-emitting devices and lasers), sensors, transistors, and photovoltaics.[1-7] All of these applications have used lattice-matched (or slightly mismatched) semiconductor systems such as Si/SiGe, GaAs/AlGaAs, or GaN/AlGaN.[8-10] While a tremendous amount of effort has been devoted to these systems to create high-quality heterojunctions, these heterogeneous integrations must be made within a very strict lattice match rule.[11-15] To overcome this limitation and restriction, namely to integrate dissimilar/lattice mismatched semiconductors, various novel approaches have been proposed. In the early day, advanced wafer bonding techniques, such as surface activated bonding (SAB) was introduced and applied to demonstrate various Si-to-III-V solar cells and transistors.[16,17] However, this method has several restrictions in that the source materials must exist in the form of the wafer, and the defect formation is due to different lattice parameters between to-be-bonded semiconductors. Recently, two heterogeneous integration methods have been reported by introducing an ultra-thin intermediate layer between two to-be-bonded semiconductors; the first method is remote epitaxy which uses mono layered graphene as a tunneling medium [18-20], and another method is the ultra-thin oxide (UO) method that uses sub-nanometer thick atomic layered deposited (ALD) oxides as a tunneling medium. [4,21,22] The significance of these two recent methods is that the high-quality arbitrary heterogeneous integration between two dissimilar single-crystalline semiconductors can be realized regardless of their crystal orientations, thermal expansion, and lattice constants. To date, most experimentally demonstrated heterostructures use graphene (for the remote epitaxy) and aluminum oxide (for the UO method); other quantum tunneling interfacial materials have not yet been explored.

In this paper, we built a numerical p-n Si/GaAs heterojunction model using quantum-mechanical tunneling theory with various quantum tunneling interfacial materials including two-dimensional semiconductors such as hexagonal boron nitride (h-BN), graphene, and ALD-enabled oxide materials such as $HfO_2$, $Al_2O_3$, and $SiO_2$. Their tunneling efficiencies and tunneling current with different thicknesses were then systematically calculated and compared. Finally, multiphysics modeling was performed with the aforementioned tunneling interfacial materials to analyze the changes in strain under different temperature conditions. Overall, our results offer a viable route to guide the selection of quantum tunneling materials for myriad possible combinations of new heterostructures that can be obtained via remote epitaxy and the UO method.

**Result and Discussion**

**Figure 1(a)-(c)** shows a schematic illustration of the heterogeneous integration of two dissimilar semiconductors sandwiched with ultra-thin oxide and 2D materials. Figure 1 shows that semiconductors A and B are different lattice-mismatched materials with an ultra-thin quantum tunneling layer. In most cases, oxides and h-BN create tunneling barriers, but the graphene layer forms a quantum well leading to different band offsets. In this calculation, we employed p-type Si and n-type GaAs; each quantum tunneling layer was seen among h-BN, graphene $HfO_2$, $Al_2O_3$, and $SiO_2$ with different thicknesses from 0.3 nm to 1.3 nm. A continuous

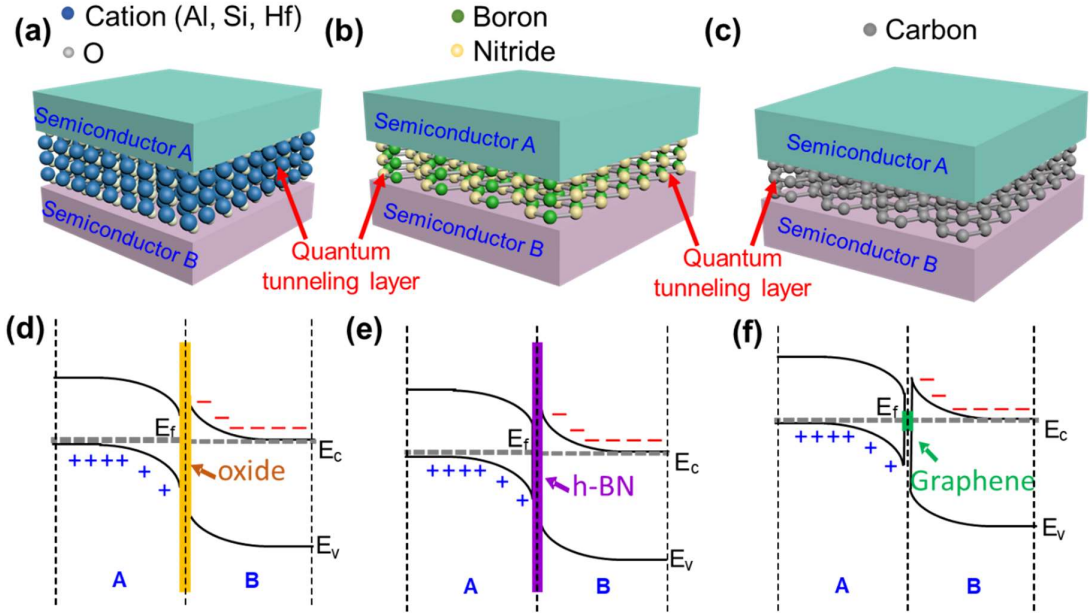

**Figure 1.** A schematic illustration of semiconductor heterostructure formed by (a) ultra-thin oxide (UO) interfacial layer such as $Al_2O_3$, $HfO_2$, $SiO_2$ (b) 2D materials such as h-BN and (c) graphene, and (d)-(f) their corresponding generic band diagrams.

wave function was employed to investigate the tunneling probability (T) of carriers (electrons and holes) across the heterointerface. A wave function incident at the tunneling barrier is a part of the wave transmitted through the barrier; another part of the wave is thus reflected back. Hence the transmitted wave is considered to be a tunneling carrier. To explain this phenomenon, we considered the one-dimensional Schrödinger wave equation as follows. A detailed solution of equation (1) for different conditions such as before the tunneling barrier (incident wave) and after the tunneling barrier (transmitted wave) was calculated.

$$\left(-\frac{\hbar^2}{2m^*}\Delta^2 + V(x)\right)\Psi(x) = E\Psi(x) \qquad (1)$$

Considering all boundary conditions, the tunneling probability is derived as the modulus squared of the transmitted wave ratio to the incident wave function. The detailed derivation of equation (1) can be found in supplementary information.

$$T = \frac{4E(V_0-E)}{V_0^2 sinh^2(k_2 a) + 4E(V_0-E)} \qquad (2)$$

where E is the potential energy of the carrier (electron or hole), $V_o$ is the potential barrier height (conduction or valance band offset), $k_2$ is related to effective mass, and $a$ is the thickness of the tunneling layer (oxide or 2D material interface). To simplify the tunneling probability across two different semiconductors, the semiconductor A-tunneling layer-semiconductor B structure (**Figure 1**) was divided into two sections: the semiconductor A-to-tunneling layer and the semiconductor B-to-tunneling layer. We calculated ten separated tunneling probability calculations because five different quantum tunneling materials were used: the tunneling probability of electrons in the $HfO_2$/n-GaAs, $Al_2O_3$/n-GaAs, and $SiO_2$/n-GaAs interfaces as well as the tunneling probability of holes in the $HfO_2$/p-Si, $Al_2O_3$/ p-Si, and $SiO_2$/p-Si interfaces. Both used equation (2). **Figure S1** in the supplementary information represents band diagrams of these structures with different oxide tunneling layers and their barrier potential for electrons and holes.

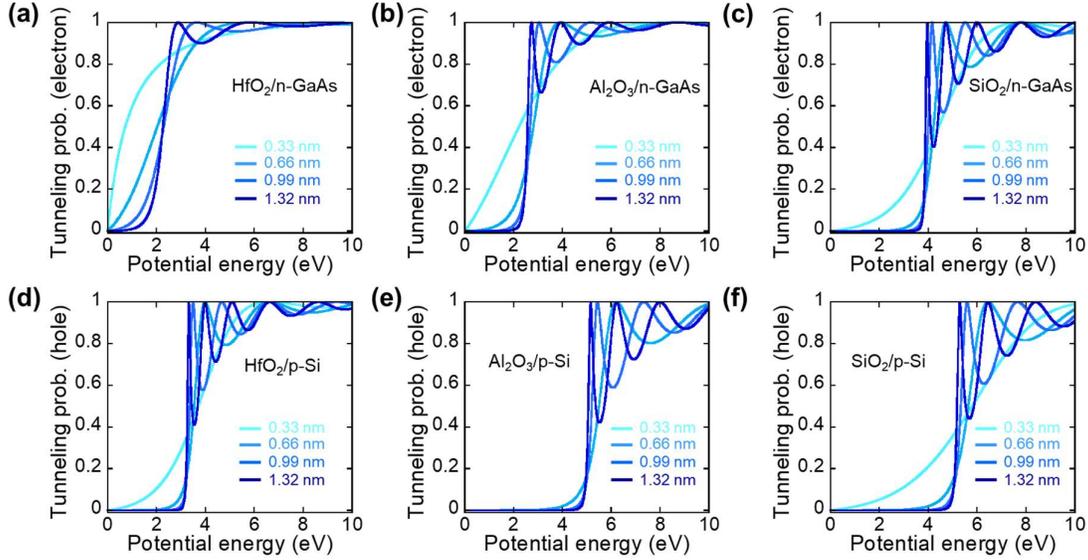

**Figure 2.** Tunneling probability of an electron across the UO interfacial layer with different thicknesses (a) $HfO_2$/n-GaAs (b) $Al_2O_3$/n-GaAs (c) $SiO_2$/n-GaAs, (d)-(f) tunneling probability of holes across these UO interfacial layers with p-Si with different thicknesses.

**Figure 2(a)-(c)** shows the tunneling probability of electrons at the n-GaAs/oxide quantum tunneling layer interface. The oxide thickness increased as 0.33 nm, 0.66 nm, 0.99 nm, and 1.32 nm, respectively. Interestingly, the effective mass for the electron and barrier height played a crucial role in determining the tunneling probability. $HfO_2$ has an electron effective mass of 0.13 $m_o$ whereas the $Al_2O_3$ and $SiO_2$ are 0.32 $m_o$ and 0.5 $m_o$, respectively. The barrier height of $HfO_2$ ($\Phi_B$=1.9eV) is lower than $Al_2O_3$ ($\Phi_B$ =2.33 eV) and $SiO_2$ ($\Phi_B$ =3.7 eV). Hence, a lighter electron effective mass leads to a lower barrier height and thus a higher tunneling probability.

In this particular example (Si/GaAs p-n junction), it is clear that the $HfO_2$ tunneling material is more efficient than the $Al_2O_3$- and $SiO_2$-based tunneling layer. In addition, tunneling in quantum mechanics occurs frequently and thus leads to increased carrier transport across the heterointerface when the potential energy (E) is smaller than the barrier height ($\Phi_B$). [23,24] Figure 2 shows that when the thickness of the oxide quantum tunneling layer increased from 0.33 nm to the 1.3 nm, the tunneling probability of electron also decreased at E<V cases. [25] When the potential energy (E) was higher than barrier height (V), the electron can roll over the barrier, and thus pass the barrier by the classical theory dominates. However, the energy quantization effect dominates when the oxide thickness gets thinner from 1.3 nm to 0.33 nm; thus, there is also a probability of electron tunneling via the E>V condition upon quantum tunneling. Quantum scattering theory describes the energy quantization effect such as the quantum well structure.[26] Electrons can pass through the barrier when the energy of an electron is greater than 1.9 eV (**Figure 2(a)**). As the thickness of HfO$_2$ gets thicker, more discrete energy levels (resonance peak) appear and suddenly spike the tunneling probability close to unity. The effective mass of the electron defines the number of resonant peaks; in other words, a higher effective mass causes more resonant peaks to appear. For example, one resonant peak appears when the Al$_2$O$_3$ thickness is 0.66 nm (**Figure 2(b)**), but two and three resonant peaks appear when the Al$_2$O$_3$ thicknesses are 0.99 nm and 1.3 nm, respectively. In addition, the SiO$_2$ case shows stronger oscillations in resonant peaks due to a heavier electron effective mass (0.5 m$_o$) (**Figure 2(c)**). Interestingly, the resonant peak becomes wider when the potential energy exceeds their corresponding barrier height (for example, >4.1 eV for SiO$_2$ in **Figure 2(c)**). This can be explained by both the tunneling and overflow of the barrier at the same time. **Figure 2(d)-(f)** represent the tunneling probability of holes at the p-Si/oxide quantum tunneling layer interface. A higher effective hole mass in HfO$_2$ (0.58m$_o$) leads to more resonant peaks than Al$_2$O$_3$- and SiO$_2$-based structures. However, HfO$_2$ shows a nearly 50% higher tunneling probability than Al$_2$O$_3$ and SiO$_2$ due to the lower barrier height (3.1 eV) relative to Al$_2$O$_3$ and SiO$_2$. The transport mechanism of the carriers has an exponential relationship with the thickness of the tunneling barrier.[27] As shown in the previous section, the tunneling probability across the barrier is reduced with increased tunneling layer thickness. Based on these relationships, we calculated the current density across the tunneling layer using different oxides at different thicknesses to investigate the carrier transport mechanism. The current density was calculated from $J = qnV_R T(E)$ where "n" is the carrier density, T(E) is the tunneling probability, and $V_R$, is the Richardson velocity for the electron and the hole. $V_R$ was achieved from $\sqrt{kT/2\pi m_*}$ where k is the Boltzmann constant, T is room temperature, and $m_*$ is the effective mass of electrons and holes. **Figure 3(a-c)** represent the electron current density at HfO$_2$/n-GaAs, Al$_2$O$_3$/n-GaAs, and SiO$_2$/n-GaAs with different oxide thicknesses, respectively. In general, the calculated current density increases as the tunneling layer gets thinner from 1.32 nm to 0.33 nm for all three oxide cases.

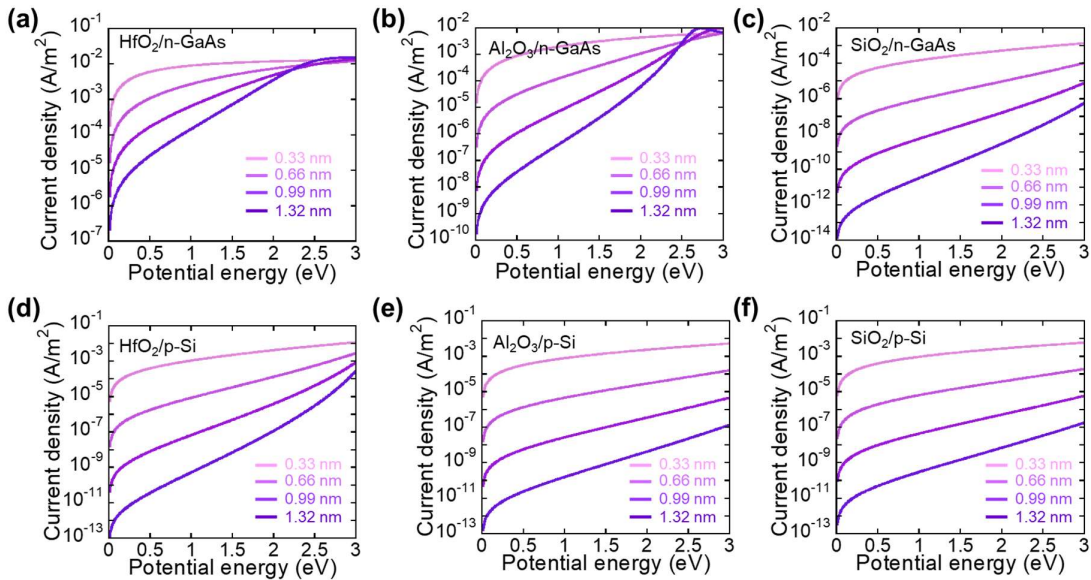

**Figure 3.** (a)-(c) Calculated current density of electrons and (d)-(f) holes with different oxide thicknesses across HfO$_2$/n-GaAs, Al$_2$O$_3$/n-GaAs, and SiO$_2$/n-GaAs.

The calculated current density for both Si and GaAs with the HfO$_2$ tunneling layer was substantially higher (at least one order of magnitude higher) than the same structures with a Al$_2$O$_3$ and SiO$_2$ tunneling layer. **Figure 3(d)-(f)** represents the hole current density at HfO$_2$/p-Si, Al$_2$O$_3$/p-Si, and SiO$_2$/p-Si with different oxide thicknesses, respectively, which show the increasing trend in the hole current density as the oxide layer gets thinner. From this result, we conclude that the ideal oxide tunneling material should have a lower band offset (barrier height), and lighter effective mass to both semiconductors A and B. In addition to these parameters, several practical aspects such as thin-film coverage in a nanoscale thickness, surface passivation capability, and controllability of native oxide removal can also affect the tunneling probability across the interface.

We further investigated two promising 2D material-based tunneling layers: The h-BN and graphene used in remote epitaxy. These two materials have completely opposite material properties including bandgap (0 eV for graphene vs. 5.97 eV for h-BN), effective mass (~0.01 m$_0$ for graphene vs. > 0.2 for h-BN), and doping (highly conductive for graphene vs. semi-insulating for h-BN).[28-30] Therefore, a comparison of the tunneling properties and tunneling current using these two materials offers material selection guideline among the numerous 2D materials for remote epitaxy. The parameters for graphene and h-BN can be found in supplementary information (**Table S1**). **Figure 4(a) and (b)** represent the hole and electron tunneling probability of h-BN/p-Si and h-BN/n-GaAs. The tunneling probability shows a similar trend with other oxide tunneling materials. This might be attributed to a wide bandgap (5.97eV) and a similar range of barrier height with oxides.[31] Interestingly, the tunneling structures using h-BN show more resonant tunneling peaks than the oxide-based tunneling structures due to its larger effective mass. **Figure 4(c) and (d)** show the hole and electron tunneling probability of graphene/p-Si and graphene/n-GaAs. Graphene has unique material properties: It is tunable and has an energy bandgap close to zero. It has the lightest effective mass, etc. The very conductive nature of monolayer graphene could be highly efficient for carrier transport across the heterostructure. Although the tunneling probability is reduced with an increased layer thickness (tri-layer), the tunneling probability remains very high due to the negative band offsets with Si and GaAs. The current density with the h-BN and graphene tunneling layer were also calculated using the same method as the oxide tunneling structures. This predicts their transport properties as shown in **Figure 5**. The calculated current density decreased with increasing graphene thickness. The reduced current density is shown in **Figure 5(a) and (b)** up to the five-layered h-BN. **Figure 5 (c) and (d)** show the current density vs. potential energy plots with the monolayer, bilayer, and trilayer graphene. The highest tunneling current density shown in the graphene case might be attributed to the smallest bandgap and lightest effective masses. However, the formation of the quantum well with the graphene case might cause problems for specific devices based on quantum barrier/wells or carrier transport along with heterointerfaces because the graphene layer can trap the carriers and lead to high interface defects.

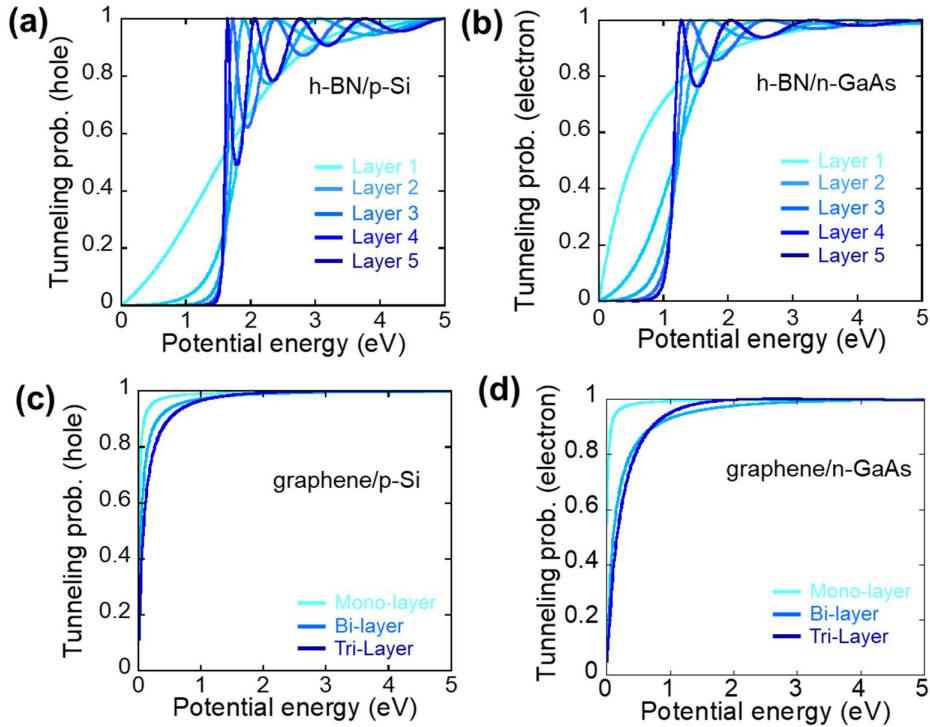

**Figure 4.** Tunneling probability of (a) hole and (b) electron across h-BN/p-Si and h-BN/n-GaAs with different h-BN thicknesses, and (c) hole and (d) electron across graphene/p-Si and graphene/n-GaAs with different graphene thicknesses.

Besides the transport property across the heterojunction, it is important to evaluate the changes in the strain of the tunneling layer under different temperature conditions. Smaller strains at the tunneling layer offer more reliable heterogeneous integration between two dissimilar semiconductors that typically have different thermal expansion coefficients. To investigate the evolution of the strain at the tunneling layer under different temperature conditions, we performed multiphysics simulations with five n-GaAs/tunneling layer/p-Si models using h-BN, graphene, HfO$_2$, Al$_2$O$_3$, and SiO$_2$ were created and traced the strain of the tunneling layer from room temperature (300 K) to 200 °C (473 K) using COMSOL$^{TM}$. **Figure 6** summarizes the strain as a function of temperature for five tunneling layers (Details of the simulation can be found in **Figure S2** of the supplementary information). While SiO$_2$, h-BN, and graphene showed almost negligible degrees of the strain changes (< 0.1 %), HfO$_2$ particularly experienced nearly ten times higher strain by heating. Given that the thermal expansion coefficient of HfO$_2$ is much larger (6.0 μm/m-K)[32] compared to other tunneling layer materials (0.5 μm/m-K or less)[33], it is possible to conclude that HfO$_2$ is not a favorable material choice for the tunneling layer because HfO$_2$ tunneling layer can cause layer delamination during the fabrication process or device operation by heating. In addition, the processability is another important point that governs the selection of the effective tunneling interfacial layer. For example, SiO$_2$ has a wider bandgap than other oxides in this study, but it has a lower dielectric constant. HfO$_2$ suffers from crystallization at high temperatures due to a fast diffusion of oxygen through the HfO$_2$. This results in uncontrolled growth of interfacial layers.[34] Therefore, considering all electrical and mechanical characteristics of various interfacial layer materials, the ideal interfacial layer should have a small conduction and valence band offsets, small effective mass, and small difference in thermal expansion coefficient to the to-be-bonded materials. Most of all, ideal interfacial materials should be formed by ALD or existed as a 2D format because they require an atomic scale thickness control for the perfect surface coverage.

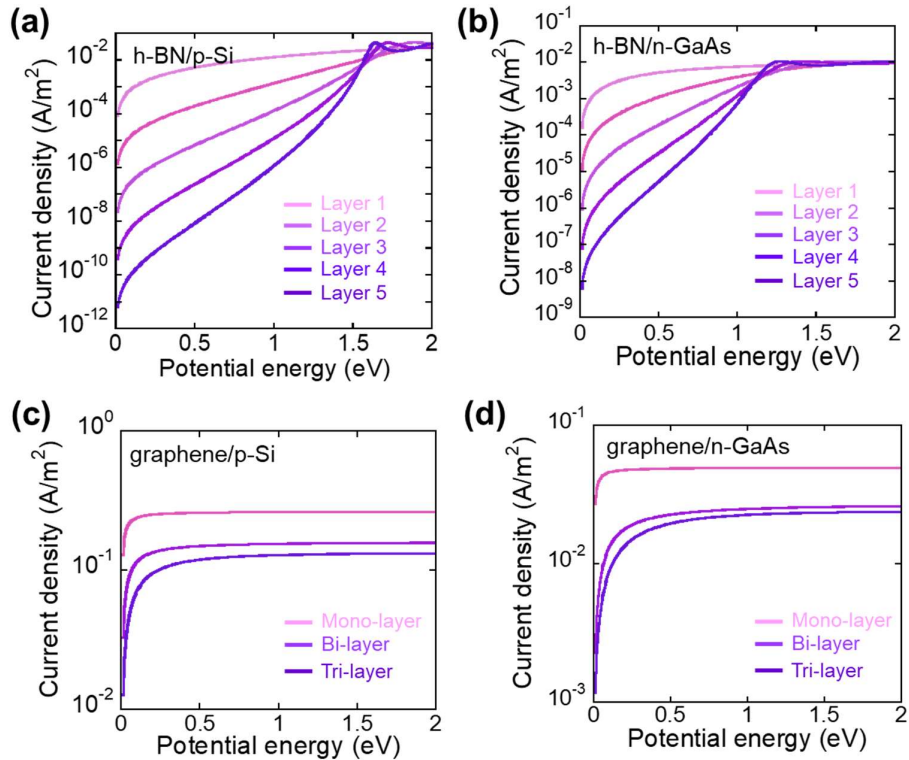

**Figure 5.** Calculated hole and electron current density across the (a) h-BN/p-Si, (b) h-BN/n-GaAs, (c) graphene/p-Si and (d) graphene/n-GaAs with different thicknesses.

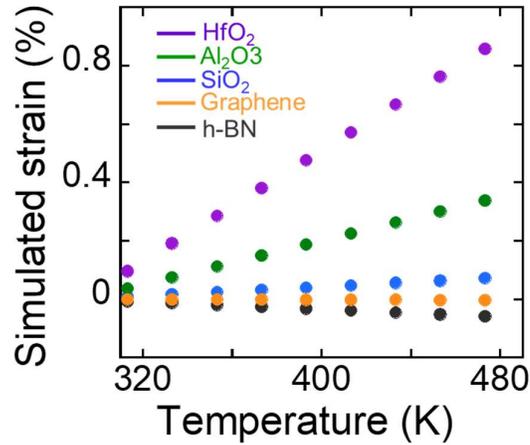

**Figure 6.** Simulated strain (%) as a function of temperature for five interfacial layers (HfO$_2$, Al$_2$O$_3$, SiO$_2$, graphene and h-BN) with different temperature ranging from 300 K to 473 K.

**Conclusions**

In conclusion, we have built a numerical p-n Si/GaAs heterojunction model using a quantum-mechanical tunneling theory with various quantum tunneling interfacial materials including two-dimensional semiconductors such as hexagonal boron nitride (h-BN) and graphene and ALD-enabled oxide materials such as HfO$_2$, Al$_2$O$_3$, and SiO$_2$. Their tunneling efficiencies and tunneling current with different thicknesses were systematically calculated and compared. Multiphysics modeling was used with the aforementioned tunneling interfacial materials to analyze changes in strain under different temperature conditions. Considering the transport properties and thermal-induced strain analysis, $Al_2O_3$ among three oxide materials and graphene in 2D materials are favorable material choices that offer the highest heterojunction quality. Overall, our results offer the viable route to guide the selection of quantum tunneling materials for myriad possible combinations of new heterostructures that can be obtained via remote epitaxy and the UO method.

**Acknowledgment**
This work was supported by the National Science Foundation (Grant number: ECCS - 1809077) and partially by the seed grant by Research and Education in energy, Environment, and Water (RENEW) Institute at the University at Buffalo.

**Supporting Information:**
Supporting information is available from the online page.

Supplementary Information

for

# Theoretical Prediction of Heterogeneous Integration of Dissimilar Semiconductor with Various Ultra-Thin Oxides and 2D Materials

*Md Nazmul Hasan[1], Chenxi Li[1], Junyu Lai[1], Jung-Hun Seo[1]\**

Department of Materials Design and Innovation, University at Buffalo, the State University of New York (SUNY), Buffalo, New York, U.S.A, 14260

*Email: junghuns@buffalo.edu

**Theoretical and mathematical background:**

The main, well-established argument of the semiconductor's tunneling mechanism is that the electron is considered a continuous wave function. When a wave function incident at the barrier, part of the wave is transmitted through the barrier, and part of the wave is reflected back. Hence, the transmitted wave is considered as tunneling electrons. Two possible cases could arise that the particle's total energy can either be higher than the potential barrier or lower than the potential barrier. If the particle energy is sufficiently high enough, it can roll over the barrier that meets classical mechanics. But if the particle energy is not sufficiently high enough, then tunneling happens.

To explain this tunneling phenomenon, we have considered the one-dimensional Schrödinger wave equation as follows-

$$\left(-\frac{\hbar^2}{2m^*}\Delta^2 + V(x)\right)\Psi(x) = E\Psi(x) \tag{1}$$

Where the $m^*$ is effective mass, $\hbar$ is plank constant, V(x) is barrier height and considered piecewise constant. It will not be straightforward if the V(x) varies with the position in the x-direction. For a thin rectangle potential barrier, the height at the Interface can have three different zones. If "a" is the barrier's thickness, the incident wave would see barrier height, V=0 before and after the transmission, but its height is considered V=Vo in the middle of the barrier. So, in three regions, the Schrödinger wave equation can be written as follows:

$$\left(-\frac{\hbar^2}{2m^*}\Delta^2 + V(x)\right)\Psi_1(x) = E\Psi_1(x), \quad \text{region 1: [x<0 and V=0]} \tag{2}$$

$$\left(-\frac{\hbar^2}{2m^*}\Delta^2 + V(x)\right)\Psi_2(x) = E\Psi_2(x), \quad \text{region 2: [x<0<x<a and V=V}_0\text{]} \tag{3}$$

$$\left(-\frac{\hbar^2}{2m^*}\Delta^2 + V(x)\right)\Psi_3(x) = E\Psi_3(x), \quad \text{region 3: [x>a and V=0]} \tag{4}$$

The solutions of equation (2-4) can be assumed as follows:

$$\Psi_1(x) = Ae^{ik_1x} + Be^{-ik_1x} \tag{5}$$
$$\Psi_2(x) = Ce^{ik_2x} + De^{-ik_2x} \tag{6}$$
$$\Psi_3(x) = Ee^{ik_3x} + Fe^{-ik_3x} \tag{7}$$

Where $k_1^2 = 2m^*E/\hbar^2$, $k_2^2 = 2m^*(E-V_0)/\hbar^2$ and $k_3^2 = \frac{2m^*E}{\hbar^2} = k_1^2$

Since there is no potential disturbance after the wave is wholly transmitted in the 3rd region, F=0, the probability of finding an electron in region 3 is constant and like to appear. After considering so many boundary conditions and solving equation (5-7), it can be written as follows:

$$\frac{B}{A} = \frac{(k_1^2 - k_2^2)(1 - e^{i2ak_2})}{(k_1+k_2)^2 - (k_1+k_2)^2 e^{i2ak_2}} \tag{8}$$

$$\frac{F}{A} = \frac{4k_1 k_2 (e^{i(k_2-k_1)a})}{(k_1+k_2)^2 - (k_1+k_2)^2 e^{i2ak_2}} \tag{9}$$

The tunneling probability is the modulus squared of the transmitted wave ratio to the incident wave function. So it can be defined as:

$$T = \left|\frac{F}{A}\right|^2 = \frac{4E(V_0-E)}{V_0^2 \sinh^2(k_2 a) + 4E(V_0-E)} \quad (10)$$

In the case of a thick barrier, the electron wave will be reflected even the total given energy is greater than the barrier height potential. Hence all-electron can't tunnel through the potential barrier. **Figure S1** shows a schematic illustration of the energy bandgap with an ultra-thin oxide tunneling layer. The conduction band offset ($\Delta E_C$) for different ultrathin oxides with n-GaAs, considered at the tunneling barrier potential for electrons. Similarly, the valence band offset ($\Delta E_V$) for various ultra-thin oxides with p-Si. The electron and hole can easily tunnel thru the oxides if the thickness gets thinner.

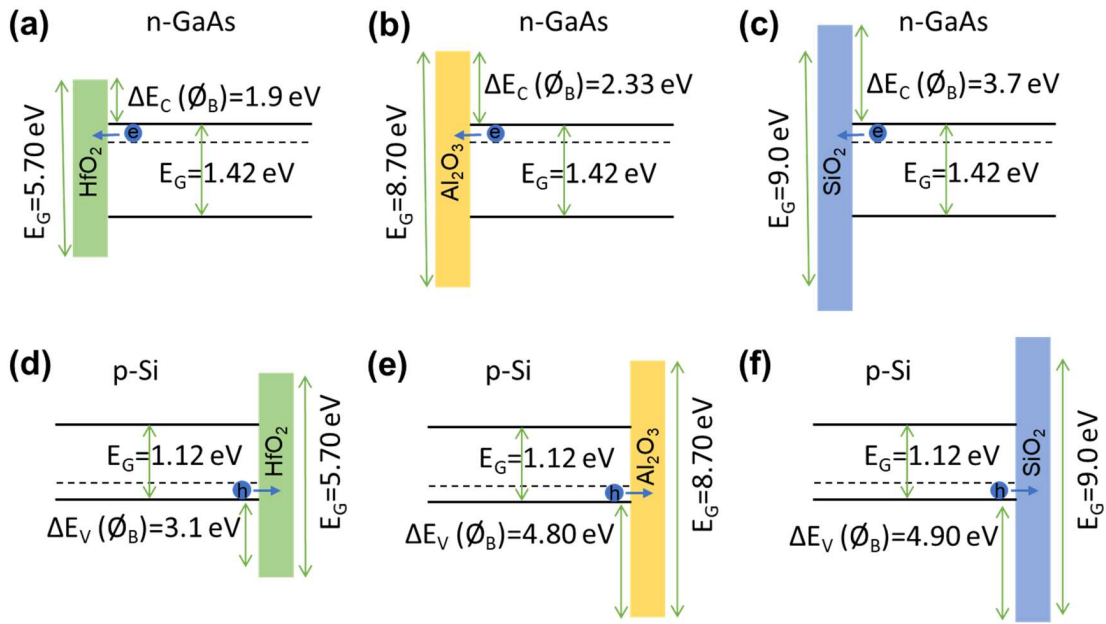

**Figure S1**: A schematic illustration of energy bandgap with (a)-(c) UO interfacial layers with n-GaAS and (d)-(f) UO interfacial layers with p-Si.

A COMSOL™ Multiphysics simulation was carried out to calculate the strain for different oxides and 2D materials induced at the heterointerface between p-Si and n-GaAs. A temperature-dependent strain calculation for different oxides ($Al_2O_3$, $HfO_2$, $SiO_2$) and 2D materials (graphene, h-BN) have been shown in **Figure S2**.

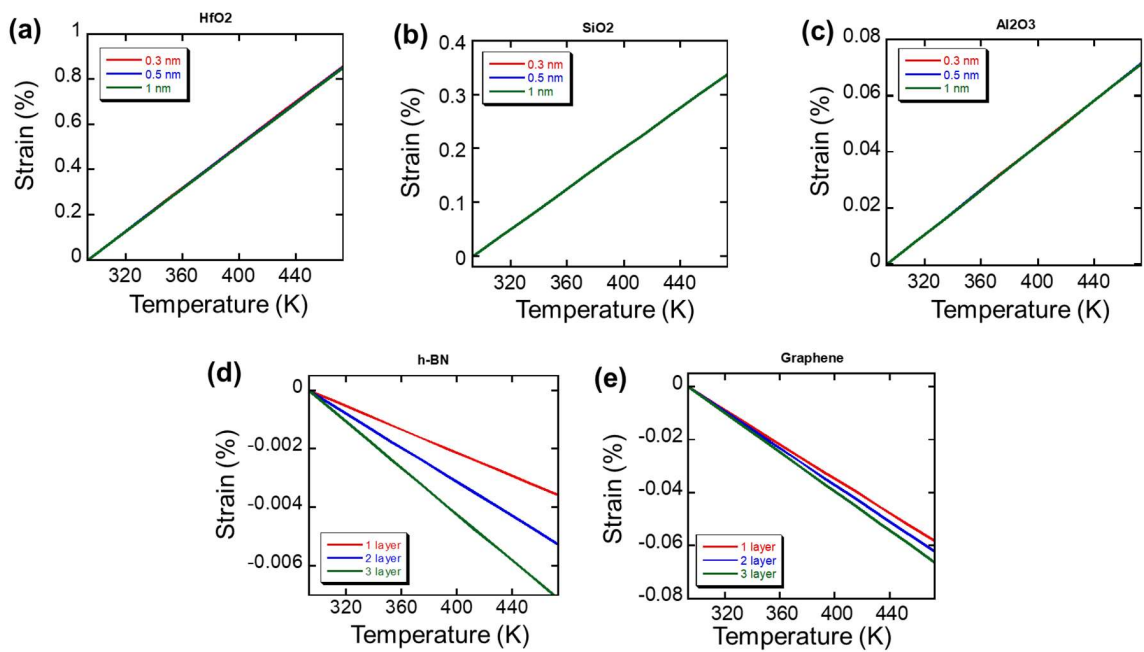

**Figure S2.** Simulated strain values of (a)-(c) UO interfacial layers and (d)-(e) 2D materials with different thicknesses as a function of temperature.

**Table S1:** The parameters for oxides (HfO$_2$, Al$_2$O$_3$, and SiO$_2$) and 2D materials (h-BN and graphene) that are used to calculate the transport property of the heterojunctions.[1-27]

|  | HfO$_2$ | Al$_2$O$_3$ | SiO$_2$ | h-BN | Graphene |
|---|---|---|---|---|---|
| **Electron Effective mass (m$_e$*)** | 0.13 m$_o$ [1] | 0.32 m$_o$ [2] | 0.5 m$_o$ [3] | 0.26 m$_o$ [4,5] | Monolayer: 0.012 m$_o$ [6]<br>Bilayer: 0.041 m$_o$ [7]<br>Trilayer: 0.052 m$_o$ [8] |
| **Hole Effective mass (m$_h$*)** | 0.58 m$_o$ [9] | 0.36 m$_o$ [10] | 0.33 m$_o$ [11] | 0.47 m$_o$ [12,13] | Monolayer: 0.013 m$_o$<br>Bilayer: 0.036 m$_o$ [14]<br>Trilayer: 0.038 m$_o$ [15] |
| **Band Gap (eV)** | 5.70 [1] | 8.70 [1] | 9.0 [1] | 5.97 [17] | Close to 0, no more than 0.3 [16] |
| **p-Si: Conduction band offset/barrier (eV)** | 1.5 [1] | 2.8 [1] | 3.1 [1] | 1.0, 1.5 [17] | Monolayer: -0.74 eV<br>Bilayer: -0.45 eV<br>Trilayer: -0.46 eV |
| **p-Si: Valence band offset/barrier (eV)** | 3.1 | 4.80 | 4.90 | 4.97, 4.47 | Monolayer: 0.74 eV<br>Bilayer: 0.45 eV<br>Trilayer: 0.46 eV |
| **n-GaAs: Conduction band offset/barrier (eV)** | 1.9 | 2.33 [18] | 3.7 [19] | 1.02 [20] | Monolayer: -0.68 eV [21]<br>Bilayer: -0.75 eV [21]<br>Trilayer: -0.78 [22] |
| **n-GaAs: Valence band offset (eV)** | 2.38 | 4.95 | 3.88 | 4.95 | Monolayer: 0.68 eV<br>Bilayer: 0.75 eV<br>Trilayer: 0.78 eV |
| **Density (g/cm$^{-3}$)** | 9.68 | 3.98 | 2.65 | 2.1 | 2.267 |
| **Thermal expansion coefficient (/K)** | 6 [23] | 4.4E$^{-6}$ [24] | 0.56 [25] | 7.2 [26] | -3.75 [27] |